\documentclass{IEEEtran}
\usepackage{lineno}
\usepackage{amssymb}
\usepackage{amsfonts}
\usepackage{amsmath}
\usepackage{algorithm}
\usepackage{algorithmic}
\usepackage{tikz}
\usepackage{graphicx,subfigure,cite,multicol,multirow,diagbox,booktabs,array}
\usepackage{color}
\usepackage{bm}
\usetikzlibrary{shapes.geometric, arrows}
\ifCLASSINFOpdf
\else
\fi

\begin{document}
\bibliographystyle{IEEEtran}
\title{Estimation of Covariance Matrix of Interference for Secure Spatial Modulation against a Malicious Full-duplex Attacker}

\author{Lili Yang, Xinyi Jiang, Feng Shu, Weibin Zhang and Jiangzhou Wang,~\IEEEmembership{Fellow,~IEEE}
\thanks{Lili Yang,~Xinyi Jiang and Weibin Zhang are with the School of Electronic and Optical Engineering, Nanjing University of Science and Technology, Nanjing, 210094, China. }
\thanks{Feng Shu is with the School of Information and Communication Engineering, Hainan University, Haikou 570228, China. and also with the School of Electronic and Optical Engineering, Nanjing University of Science and
Technology, Nanjing 210094, China. (Corresponding authors: Feng Shu). }
\thanks{Jiangzhou Wang is with the School of Engineering and Digital Arts, University of Kent, Canterbury CT2 7NT, U.K. (e-mail: j.z.wang@kent.ac.uk).}}
\maketitle

\begin{abstract}
In a secure spatial modulation with a malicious full-duplex attacker,  how to obtain the interference space or channel state information (CSI) is very important for Bob to cancel or reduce the interference from Mallory.  In this paper, different from existing work with a perfect CSI, the covariance matrix of malicious interference (CMMI) from Mallory is estimated and is used to construct the null-space of interference (NSI). Finally, the receive beamformer at Bob is designed to remove the malicious interference using the NSI. To improve the estimation accuracy, a rank detector relying on Akaike information criterion (AIC) is derived. To achieve a high-precision CMMI estimation,  two methods are proposed as follows: principal component analysis-eigenvalue decomposition  (PCA-EVD), and joint diagonalization (JD).
The proposed PCA-EVD is a rank deduction method whereas the JD method is a joint optimization method with improved performance in low signal to interference plus noise ratio (SINR) region at the expense of increased complexities. Simulation results show that the proposed PCA-EVD performs much better than the existing method like sample estimated covariance matrix (SCM) and EVD in terms of normalized mean square error (NMSE) and secrecy rate (SR). Additionally, the proposed JD method has an excellent NMSE performance better than PCA-EVD in the low SINR region (SINR$\leq $0dB) while in the high SINR region PCA-EVD performs better than JD.
\end{abstract}


\IEEEpeerreviewmaketitle
\section{Introduction}
Wireless communications have been developed rapidly in recent years \cite{8718991}. Spatial Modulation (SM) emerges as a novel multiple-input-multiple-output (MIMO) transmission in wireless communication, which combines both antenna indexes and modulation symbols in the signal constellation to transmit information\cite{4382913}. Unlike conventional transmit methods, this technique can efficiently avoid inter-channel interference (ICI) and transmit antennas synchronizations (TAS) as only one antenna is activated during one transmission slot. By reducing the radio frequency (RF) links, SM can obtain higher energy efficiency (EE) than Bell Labs layered space-time (BLAST) and space-time coding (STC). In SM transmission, a block of information bits is mapped into two parts: a symbol chosen from \emph{M}-ary constellation diagram and an index of an activated transmit antenna.
In the way of combining these two units, SM can achieve better spectral efficiency (SE) with the number of transmit antennas being the power of two.

As the transmit environment for SM is an open space, it is very likely for eavesdroppers to capture confidential message (CM)\cite{8911437}. In traditional ways, encryption technology is used to guarantee information away from being acquired by eavesdroppers. But it requires careful key distribution and service management \cite{708447}. Nowadays, physical-layer security has attracted wide interest from information-theoretical perspective \cite{8611204}. In \cite{6187751}, authors proposed a linear precoding method against wiretap channels to improve the secrecy rate by utilizing the important relationship between MIMO mutual information and the received minimum mean square error (MSE). To reduce the information that eavesdroppers can obtain, artificial noise (AN) was introduced in \cite{9262009,7116516} to interfere with eavesdroppers by projecting it onto the null-space of the desired channel. The authors investigated transmit antenna selection (TAS) schemes in \cite{8373751} to enhance the secrecy rate performance of the SM system.

As mentioned above, since eavesdroppers may acquire the CM and are passive, in\cite{7740338}, the authors proposed a full-duplex (FD) eavesdropper that can not only overhear CM but also send the jamming signal to the legal receiver to interfere the CM. Here, the eavesdropper became an active malicious attacker. And the eavesdropper projected the jamming signal onto the null-space of its CM receiving channel, therefore it will not suffer from its jamming. In \cite{9116808}, the authors discussed the impact of an FD attacker on system performance. Several efficient beamforming methods against the malicious jamming at receiver have been designed in \cite{9325946}.

In the above investigations, the perfect channel state information (CSI) between attacker and receiver was assumed as a prior condition. However, it is impossible to obtain a perfect CSI in practice. Actually, only imperfect CSI can be attained. In what follows, we will show how to obtain CSI. Instead of CSI, we will estimate the covariance matrix of malicious interference (CMMI) from Mallory. Once CMMI is gotten, the corresponding receive beamforming vectors is readily derived. Our main contributions are summarized as follows:


\begin{enumerate}
    \item To accurately estimate the CMMI and efficiently reduce the malicious jamming from Mallory, a rank detector of  CMMI is proposed using Akaike information criterion (AIC). The estimated rank is used to improve the precision of estimating CMMI. With the knowledge of rank assisted, the principal component analysis-eigenvalue decomposition (PCA-EVD) method is proposed to estimate the CMMI from rank reducing perspective.
    The corresponding  Cramer-Rao Lower Bound (CRLB) is also derived  as a metric to evaluate the normalized mean square error (NMSE)  performance.
    The simulation results show that the proposed PCA-EVD outperforms conventional methods EVD and sample covariance matrix (SCM) in terms of NMSE.

	\item  To reduce the performance gap between PCA-EVD and CRLB,  instead of dealing with received signals directly, a joint diagonalization (JD) is proposed to process the SCMs in a parallel way. With a unitary constraint, the proposed JD estimates CMMI by calculating the unitary matrix and diagonal matrix separately. This may achieve a substantial performance improvement.
	In accordance with simulation, in terms of NMSE, the proposed two schemes perform better than SCM and EVD in the low signal to interference plus noise ratio (SINR) region.  In particular, when SINR$\leq$-5dB, the performances of both the proposed JD and the CRLB are close as SINR decreases.
\end{enumerate}

\emph{Organization:} Section \uppercase\expandafter{\romannumeral2} describes the system model of the classical secure SM system. In Section \uppercase\expandafter{\romannumeral3}, we introduce the rank detection strategy and two covariance matrix estimation schemes are proposed. Performance simulation and analysis for the proposed methods  are presented in Section \uppercase\expandafter{\romannumeral4}  and we draw our conclusion in Section \uppercase\expandafter{\romannumeral5}.

\emph{Notations:} Boldface lower case and upper case letters denote vectors and matrices, respectively. Scalars are denoted by the lower case, $(\cdot)^{H}$ denotes conjugate and transpose operation. $||\cdot||_F$ denotes F-norm. $\mathbb E[\cdot]$ represents the expectation operation and $t_{ij}$ represents the $\emph{ij}$th element in $\mathbf T(i,j)$. $\text{diag}(\cdot)$ denotes forming a diagonal matrix using diagonal elements of a matrix.
\section{System Model}
 \begin{figure}[htb]
	\centering
	\includegraphics[height=4.5cm,width=7.5cm]{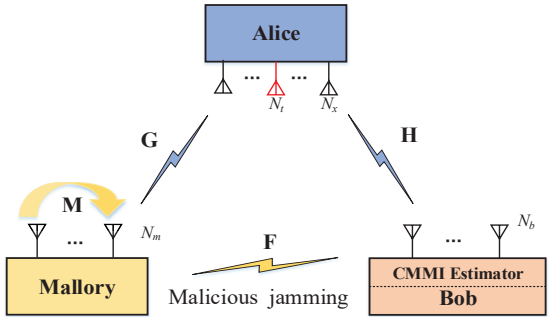}
	\caption{Block diagram model for SM.}
	\label{SM}
\end{figure}
This SM system consists of a transmitter (Alice) with $N_x$ transmit antennas, a desired receiver (Bob) with $N_b$ antennas and a full-duplex eavesdropper receiver (Mallory) with $N_m$ antennas that can not only send malicious jamming towards Bob but also capture the CM from Alice. As mentioned above, the number of activated antennas should be the power of two, therefore $N_t$ activated antennas are selected from $N_x$ transmit antennas and $N_t$ is equal to $2^{[\lg_2N_x]}$. Fig.~\ref{SM} exhibits the system model of secure SM. Referring to the secure SM model in \cite{9325946}, the transmit signal from Alice with the aid of AN and the malicious jamming from Mallory can be represented respectively as:
\begin{equation}
	\mathbf x_a=\sqrt{\beta P}\mathbf e_ns_m+\sqrt{(1-\beta)P}\mathbf{T}_{AN}\mathbf n_A,
\end{equation}
\begin{equation}
	\mathbf x_m=\sqrt{ P_M}\mathbf{P}_{J}\mathbf n_m,
\end{equation}
where $\beta$ denotes the power allocation (PA) factor, \emph P and $\emph P_M$ denote the transmit power of confidential signal  and jamming transmit power. In (1), $\mathbf e_n$ is the $n$th column of an identity matrix $\mathbf I_{Nt}$ implying that the $n$th antenna is chosen to transmit symbol where $n\in \{1,2,\dots ,N_t\}$, $s_m$ is the $m$th input symbol from the \emph{M}-ary signal constellation,  where $m\in \{1,2\dots M\}$, and $\mathbf T_{AN} \in \mathbb{C}^{N_t\times N_t}$ represents the projected AN matrix, $\mathbf n_A\sim {\mathbb{CN} (0,\sigma_A^2\mathbf I_{Nt})}$ is the random AN vector. $\mathbf P_J\in \mathbb{C}^{N_m\times N_m'}$, where $N_m>N_m'$, is the transmit beamforming matrix of jamming vector $\mathbf n_m\sim {\mathbb{CN} (0,\sigma_m^2\mathbf I_{N_m'})}$.

Therefore, the signals observed at Bob and Mallory can be expressed as follows:
\begin{align}
		 \mathbf y_b={}& \sqrt{\beta P}\mathbf{H}\mathbf{S}\mathbf e_ns_m+  \\
\nonumber		& \sqrt{(1-\beta)P}\mathbf{H}\mathbf{S}\mathbf{T}_{AN}\mathbf n_A+(\underbrace{\sqrt{P_M}\mathbf{F}\mathbf{P}_{J}\mathbf n_m+\mathbf n_B}_{\mathbf y_J}),
\end{align}
\vspace{-0.8cm}
\begin{align}
     \mathbf y_m={}& \sqrt{\beta P}\mathbf{G}\mathbf{S}\mathbf e_ns_m+ \\
\nonumber &\sqrt{(1-\beta)P}\mathbf{G}\mathbf{S}\mathbf{T}_{AN}\mathbf n_A+\sqrt{P_M}\mathbf{M}\mathbf{P}_{J}\mathbf n_m+\mathbf n_M,
\end{align}
where $\mathbf H \in \mathbb{C}^{N_b\times N_x}$, $\mathbf{G} \in \mathbb{C}^{N_m\times N_t}$, $\mathbf{F}\in \mathbb{C}^{N_b\times N_m}$ and $\mathbf{M}\in \mathbb{C}^{N_m\times N_m}$ are the channel gain matrices from Alice to Bob,  Alice to Mallory,  Mallory to Bob,  and Mallory's self-interfere channel respectively. Besides, $\mathbf S$ is the activated  antennas selection matrix, $\mathbf {S} \in \mathbb{R}^{N_x\times N_t}$. In addition, $\mathbf n_B\sim {\mathbb{CN} (0,\sigma ^2_B\mathbf I_{N_b})}$ and $\mathbf n_M\sim {\mathbb{CN} (0,\sigma ^2_M\mathbf I_{N_m})}$ represent the receiver complex additive white Gaussian noise (AWGN) vectors at Bob and Eve respectively.

The CSI from Mallory to Bob is often assumed as a perfect condition, while it is hard to obtain in practice, therefore it is necessary to estimate it to reduce the impact of malicious interference. Here, we assume there are two time slots, in the first slot, Alice ceases working and only Mallory emits jamming signals, then Bob estimates the CMMI. In the second slot, Alice and Mallory transmit signals simultaneously and Bob utilizes the estimated CMMI to remove the impact of interference from Mallory.

The malicious interference signal plus receiver noise received at Bob in the first time slot can be represented as,
\begin{equation}
	\mathbf y_{Jk}=\sqrt{P_M}\mathbf{F}\mathbf{P}_{J}\mathbf n_m+\mathbf n_B.
\end{equation}
Since the interference and noise are independent, the covariance matrix $\mathbf{R}_{y y}$ for $\mathbf y_{Jk}$ is given as:
\begin{align}\label{R-yy}
	\nonumber	\mathbf{R}_{y y}=&\mathbb E[\mathbf y_{Jk}\mathbf y_{Jk}^H]
	=P_M\sigma_m^2\mathbf{F}\mathbf{P}_J\mathbf{P}_J^H\mathbf{F}^H+\mathbf{n}_B\mathbf{n}_B^H\notag\\
	=&\mathbf{R}_{JJ}+\sigma _B^2\mathbf I_{N_b} \nonumber\\
	=&\left[\mathbf{E}_{U} \mathbf{E}_{N}\right] \mathbf \Sigma\left[\mathbf{E}_{U} \mathbf{E}_{N}\right]^{H}
\end{align}
where  $\mathbf{R}_{JJ}$ is the covariance matrix of receive jamming signal from Mallory at Bob,  $\mathbf{E}_{U}=[\mathbf u_1, \cdots, \mathbf u_r]$, $\mathbf{E}_{N}=[\mathbf u_{r+1}, \cdots, \mathbf u_{N_b}]$, and $\mathbf u_i$ denotes eigenvector corresponding to the $i$th eigenvalue ,$\lambda_i$, of CMMI $\mathbf{R}_{JJ}$, and $\mathbf \Sigma=\text {diag}[\lambda_1+\sigma_B^2,\cdots, \lambda_r+\sigma_B^2,\sigma_B^2,\cdots,\sigma_B^2]$. Here, $\lambda_1\geq \lambda_2\geq\cdots\lambda_r$, where $r$ is the rank of matrix $\mathbf{R}_{JJ}$. Actually, $\mathbf{R}_{JJ}=\sum_{i=1}^r \lambda_i\mathbf u_i\mathbf u_i^H$.

A classical assumption is to use a set of received signals $\mathbf y_{Jk}$, where $\mathbf y_{Jk}\sim {\mathbb{CN} (0, {\mathbf R_{SCM}})}$ and $k\in \{1,2\dots L\}$, where $\mathbf R_{SCM}$ represents SCM.
Based on the maximum likelihood estimation (MLE) principle and mutually independence between received signals, the SCM can be expressed as:
\vspace{-0.1cm}
\begin{equation}
	{\mathbf R_{SCM}}=\frac{1}{L}\sum_{k=1}^L \mathbf y_{Jk} \mathbf y_{Jk}^H.
\end{equation}

Given the CMMI, we can construct a receive beamforming vector of reducing the impact of the jamming from Mallory on Bob. To completely remove the jamming from Mallory, a maximized signal-to-jamming-plus-noise ratio (SJNR) can be casted as
\begin{equation}
	\mathrm {max}   \quad \mathbf{SJNR}(\mathbf{u}_{b r})
	~~\mathrm{s.t.}~~ \mathbf{u}_{br}^H\mathbf{R_{JJ}}=0_{},~~\mathbf{u}_{br}^{H} \mathbf{u}_{br}=1 .
\end{equation}
where
\begin{align}
	\nonumber   \textbf{SJNR}={}&\frac{\beta P||\mathbf{u}_{br}^H\mathbf{HS}\textbf e_ns_m||^2}{P_M\sigma_m^2||\mathbf{u}_{br}^H\mathbf{FP_{JM}}||^2+\mathbf{\sigma}_B^2\mathbf{u}_{br}^H\mathbf{I}_{Nr}\mathbf{u}_{br}}\\
	=&~~~~~~~\frac{\beta P\mathbf{u}_{br}^H\mathbf{HS}\mathbf{S}^H\mathbf{H}^H\mathbf{u}_{br}}
	{\mathbf{u}_{br}^H(\mathbf{R}_{JJ}+\mathbf{\sigma}_B^2\mathbf{I}_{Nr})\mathbf{u}_{br}}.
\end{align}

\section{Proposed Two CMMI Estimators}
In this section,  a rank detection strategy using AIC is proposed to assist estimation of the following CMMI. With the help of the estimated rank, two CMMI estimators are presented as follows: PCA-EVD, and JD. Finally, we also make a complexity comparison among them.

\subsection{Proposed rank detector}
 The rank of jamming covariance matrix $\mathbf{R}_{JJ}$ is unknown to Bob, hence it is necessary to for Bob to infer this rank. In what follows,  a rank detection strategy based on AIC is proposed. AIC is a model selection strategy based on the maximum point at empirical log-likelihood function (LLF)  \cite{2004Model}. Assuming the rank of $\mathbf{R}_{JJ}$ is $k$, $k\in \{1,2,\dots ,N_m-1\}$, the SCM has an ideal form as follows:
 \vspace{-0.2cm}
\begin{equation}
\mathbf R^{(k)}=\sum_{i=1}^{k}(\lambda_i-{\sigma^2})\mathbf{u}_i \mathbf{u}_i^H+{\sigma^2}\mathbf I_{N_b},
\end{equation}
where
\begin{equation}
    \sigma^2=\frac{\sum_{i=k+1}^{N_b}\lambda_i}{N_b-k}.
\end{equation}

To estimate the rank, the parameter vector can be represented as:
$\Theta^{(k)}=(\lambda_1,\lambda_2,\cdots,\lambda_k,\sigma^2)$.
Using $N$ samples to do the estimation, the likelihood function can be written as:
\begin{align}
    &\emph f(\mathbf{y}_{J1},\mathbf{y}_{J2},\cdots,\mathbf{y}_{Jn}|\Theta^{(k)})\\
 \nonumber   &=\prod_{i=1}^{N}\frac{1}{\pi |\mathbf R^{(k)}|}\rm{exp}-\mathbf{y}_{Ji}^H[\mathbf R^{(k)}]^{-1}\mathbf{y}_{Ji}.
\end{align}
Taking the logarithmic form of it, the LLF yields
\begin{equation}
   {\rm {ln}} \emph{L}(\Theta^{(k)})=-N\rm{ln} |\mathbf {R}^{(k)}|-tr(\mathbf{Y}_{SCM}[\mathbf {R}^{(k)}]^{-1}),
\end{equation}
where
\begin{equation}
    \mathbf{Y}_{SCM}=\frac{1}{N}\sum_{k=1}^N \mathbf {y}_{Ji} \mathbf {y}_{Ji}^H,
\end{equation}
whose eigenvalues  is utilized to estimate the  eigenvalues of $\mathbf R^{(k)}$, and  the corresponding  LLF by omitting the irrelevant items is given by
\vspace{-0.3cm}
\begin{equation}
    {\rm {ln}} \emph{L}(\Theta^{(k)})= -N\rm{ln} (\prod_{\emph i=1}^{\emph k}\widehat\lambda_i\prod_{\emph i=\emph k+1}^{\emph N_b}{\widehat{\sigma^2}}).
\end{equation}
which forms the following AIC rule for rank estimation as follows:
\begin{align}
\nonumber   \textbf{AIC}(k)=&-2{\rm {ln}} \emph{L}(\Theta^{(k)})+2K\\
          =&2N\rm{ln} (\prod_{\emph i=1}^{\emph k}\widehat\lambda_i\prod_{\emph i=\emph k+1}^{\emph N_b}{\widehat{\sigma^2}})+2(\emph k+1)
\end{align}
where the bias-correction term $\emph K$ in AIC equals the number of estimable parameters,   $K=k+1$.
The number $\emph k$ of minimizing $\textbf{AIC}(k)$ is chosen as the rank.

\subsection{Proposed PCA-EVD estimator}
As the rank $r$ for $\mathbf R_{JJ}$ is acknowledged, to estimate the CMMI efficiently, a dimension deduction strategy is utilized using the estimated rank. The main idea for a dimension deduction is the PCA \cite{2017Matrix}, where $r$ principal components are used to represent $L$ data variables that are statistical correlation, where $r$ $<$ $L$.
According to the steps of PCA, we first form a data set matrix $\mathbf X\in \mathbb{C}^{N_b\times L}$, where $\mathbf X= [\mathbf y_{J1},\cdots, \mathbf y_{JL}]$.  The covariance matrix for data is given by:
\begin{equation}
	\mathbf R_{X}=\frac{1}{P}\mathbf X\mathbf X^{H}.
\end{equation}
whose EVD produces its eigenvalues $\lambda_1, \lambda_2,\cdots,\lambda_{N_b}$ in a decreasing order similar to (\ref{R-yy}), $\mathbf{u}_1,\mathbf{u}_2,\cdots,\mathbf{u}_{N_b}$ are the corresponding eigenvectors. To cancel the channel noise and refer to (\ref{R-yy}), we first estimate the noise variance as follows

\begin{equation}
	\widehat{\sigma_{B}^2} =\frac{\lambda_{r+1}+\cdots+\lambda_{N_b}}{N_b-r}.
\end{equation}
which gives the following PCA-EVD reconstruction method
\begin{equation}
	\widehat {\mathbf{R}}_{PCA-EVD}=\sum_{i=1}^r(\lambda_i-\frac{\lambda_{r+1}+\lambda_{r+2}+\cdots+\lambda_N}{N_b-r})\mathbf u_i\mathbf u_i^H.
\end{equation}

\subsection{Proposed joint diagonalization estimator}
With the inferred rank $r$ of $\mathbf R_{JJ}$ known as a prior condition, to reduce the distance between estimated CMMI and the SCM with an aim to improve the estimated performance in the low SINR region, a JD method is proposed to estimate the CMMI by jointly optimizing SCMs below. As the estimated CMMI can be written as ${\mathbf R}_{JD}=\mathbf {T\Lambda T^H}$, where $\mathbf T$ is an unitary matrix and $\mathbf \Lambda$ is a diagonal matrix with $r$ largest  elements being nonzero,
the JD problem can be formulated as follows:
\begin{equation}\label{JD1}
	\underset{\mathbf{T,\Lambda}}{\mathrm {min}}~ \sum_{k=r+1}^L||\mathbf{T}^{H}\mathbf R_k\mathbf{T}-\mathbf \Lambda||^2_{F}  \quad
    \mathrm{s.t.}  \mathbf{T}^{H}\mathbf{T}=\mathbf{I}
\end{equation}
where
\begin{equation}
    \mathbf R_k=\frac{1}{k}\mathbf \sum_{i=1}^k\mathbf{y}_{Ji}\mathbf{y}_{Ji}^{H}-\widehat {\sigma_{B}^2}\mathbf{I}.
\end{equation}
Let us define a function $f(\cdot)$,
\begin{equation}\label{f-func}
	f(\mathbf{R}_k)=||\mathbf{R}_k-\text{diag}(\mathbf{R}_k)||^2_{F}.
\end{equation}

In the first step, provided $\mathbf \Lambda$ is taken to be $\text{diag}(\mathbf{R}_L)$, the unitary matrix $\mathbf T$ is to be calculated and the problem (\ref{JD1}) can be converted to (\ref{JD2}) by function $f(\cdot)$ in (\ref{f-func}) as follows
\begin{equation}\label{JD2}
	\underset{\mathbf{T}}{\mathrm {min}}   \quad \mathbf J=\sum_{k=r+1}^L f(\mathbf T\mathbf{R}_k\mathbf T^H),
\end{equation}
where
\begin{equation}\label{JD10}
	\mathbf T=\prod_{(i,j)\in S_{P}} \mathbf N(i,j),
\end{equation}
where $S_{P}=\{(i,j)|1\leq i< j\leq N_b\}$, matrix $\mathbf T$ is usually a unitary rotation matrix and it is obtained by multiplying $\frac{(N_b-1)^2}{2}$ Givens rotation matrices $\mathbf N(i,j)$. Referring to \cite{bookname1},  the Givens matrix $\mathbf N(i,j)$ can be regarded as an identity matrix except for the following four elements satisfying the following definition
\begin{equation}\label{JD11}
	\begin{aligned}
		\left[\begin{array}{cc}
			t_{ii} & t_{ij} \\
			t_{ji} & t_{jj}
		\end{array}\right]=\left[\begin{array}{ccccc}
			c & s \\
            -s & c
		\end{array}\right],
	\end{aligned}\quad |c|^2+|s|^2=1.
\end{equation}

As $\mathbf T$ is the product result of $\mathbf N(i,j)$, the following part describes the calculation of $\mathbf N(i,j)$. Let us define a new matrix $\mathbf R_k^{'}=\mathbf N(i,j)\mathbf R_k\mathbf N(i,j)^{H}$. Considering $\mathbf N(i,j)$ is a unitary matrix, we have $||\mathbf R_k^{'}||^2_F=||\mathbf R_k||^2_F$, combined with the definition of function $f(\cdot)$, we can get
\begin{equation}\label{JD4}
	f(\mathbf R_k^{'})+|r^{'}_{ii}|^2+|r^{'}_{jj}|^2=f(\mathbf R_k)+|r_{ii}|^2+|r_{jj}|^2.
\end{equation}
Then, $\mathbf T$ can be replaced with $\mathbf N(i,j)$ in (\ref{JD2}) and using (\ref{JD4}), the objective function (\ref{JD2}) reduces to
\begin{equation}\label{JD5}
	\mathrm {max}   \quad |r^{'}_{ii}-r^{'}_{jj}|^2,
\end{equation}
Let us define the following vectors
\begin{equation} \label{JD7}
  \mathbf v(c,s)=[|c|^2-|s|^2,c^{*}s+cs^{*}],
\end{equation}
\begin{equation} \label{JD8}
	\mathbf	g(\mathbf R_k)^H=[r_{ii}-r_{jj},r_{ij}+r_{ji}],
\end{equation}
which gives
\begin{align}
	\nonumber 	r^{'}_{ii}-r^{'}_{jj}&=(|c|^2-|s|^2)(r_{ii}-r_{jj})+(c^{*}s+cs^{*})(r_{ij}+r_{ji} ) \\
	&=\mathbf	g(\mathbf R_k)^H\mathbf v(c,s),
\end{align}
which yields
\begin{align}\label{JD3}
|r^{'}_{ii}-r^{'}_{jj}|^2=\mathbf v^H\mathbf g(\mathbf R_k)\mathbf g(\mathbf R_k)^H\mathbf v.
\end{align}
Finally, the minimization problem in (\ref{JD2}) is recasted as the maximization problem
\begin{equation}\label{JD6}
	\mathrm {max} \quad  \mathbf v^H\mathbf G\mathbf v,
\end{equation}
where
\begin{equation}
  \mathbf G=\frac{1}{L-r}\sum_{k=r+1}^{L}\mathbf g(\mathbf R_k)\mathbf g(\mathbf R_k)^{H}
\end{equation}
which, via the Rayleigh-Ritz ratio theorem,  outputs the optimal solution $\mathbf v=[x,y]^T$,  i.e., the eigenvector corresponding to the largest eigenvalue of $\mathbf G$. Substituting  $\mathbf v=[x,y]^T$ in (\ref{JD7}), the elements in the rotation matrix $\mathbf N(i,j)$ can be derived as follows:
\begin{equation}\label{JD9}
	\begin{aligned}
		c=\sqrt{\frac{x+1}{2}},
		\quad 	s=\sqrt{\frac{y^2}{2(x+1)}}.
	\end{aligned}
\end{equation}

By repeating the process from (\ref{JD7}) to (\ref{JD9}) for each pair $(i,j)$ in $S_P$,   the corresponding $\mathbf N(i,j)$ is computed, and   $\mathbf T$ is obtained through (\ref{JD10}).

Now, we turn to the second step of computing matrix $\mathbf \Lambda$. First, all diagonal elements of matrix $ \mathbf {T}^H\mathbf R_{L}\mathbf T$ are arranged in a descending order,  and its first r elements are assigned to those of $ \mathbf \Lambda$.

Finally, the estimated matrix is constructed as follows: ${\mathbf R}_{JD}=\mathbf {T\Lambda T^H}$. This completes the estimate process of ${\mathbf R}_{JD}$.

\subsection{Complexity Analysis}
To make a computational complexity comparison among the proposed schemes, we take the number of floating-point operations (FLOPs) as a performance metric.
The complexities of these methods are detailed as follows:
$\emph{C}_{SCM}\!=6KN_{r}^2$ FLOPs,
$\emph{C}_{PCA-EVD}\!=126N_{r}^3+(8K+6r-2)N_{r}^2$ FLOPs and $\emph{C}_{JD}\!=(158+8K-8r)N_{r}^3+\frac{(N_b-1)^2}{2}(126*2^3+24(K-r))+(8r+4K-8)N_{r}^2$.
Obviously, the proposed JD is at least one order of magnitude higher than PCA-EVD in terms of computational complexity from simulation.  SCM is the lowest one among them. In summary, their computational complexities have an ascending order as follows: SCM, PCA-EVD, JD.

\section{SIMULATION AND DISCUSSION}
In this section, numerical simulations are presented to analyze and compare the performances among these estimation strategies from perspectives of NMSE and secrecy rate. Simulation parameters are set as follows: $\emph{N}_b = 8$, $\emph{N}_t = 16$, $\emph{N}_m = 6$, $\emph P = 10$W, $\sigma_{br}^2 = \sigma_{mr}^2$, and quadrature phase shift keying (QPSK) is employed.
 \begin{figure}[htb]
	\centering
    \includegraphics[height=6.5cm,width=0.9\textwidth]{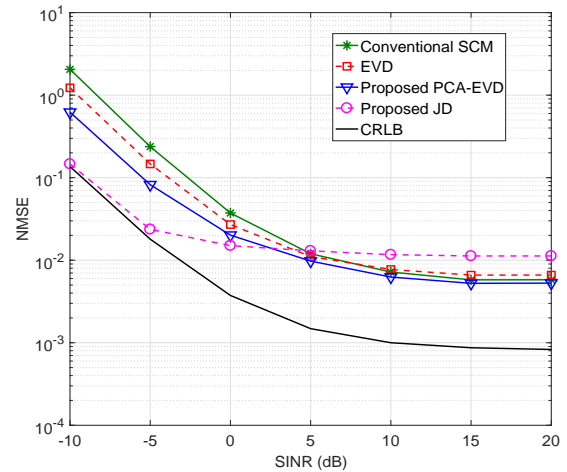}
	\caption{Comparison of NMSE versus SINR for proposed estimation methods when 8 samples are used.}
		  \setlength{\belowcaptionskip}{-1cm}
 \setlength{\belowcaptionskip}{-1cm}
	\label{NMSE}
\end{figure}

Fig.~\ref{NMSE} demonstrates the NMSE versus SINR for the proposed methods with CRLB as a performance benchmark.
From this figure, it is clearly seen that in the low SINR region, i.e., SINR$\leq$2.5dB, the proposed JD method is  the closest to CRLB compared with other methods including PCA-EVD, SCM and EVD. Besides, the descent rate of JD is slowing down with the SINR. All these methods  converge to a constant value with increase in SINR.

\begin{figure}[htb]
	\centering
	\includegraphics[height=6.5cm,width=0.8\textwidth]{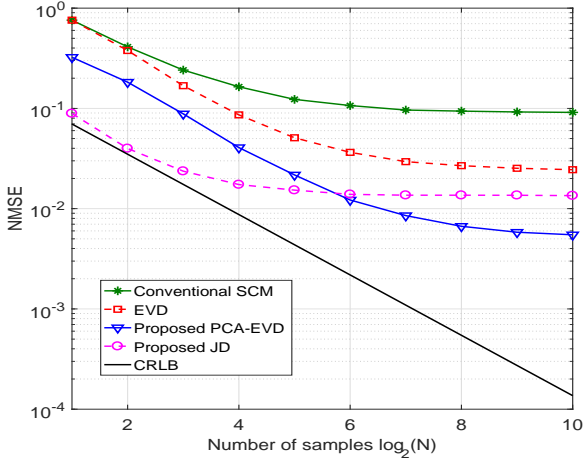}
 \vspace{-0.3cm}
	\caption{NMSE curves versus number of samples at SINR=-5$dB$.}
     \setlength{\belowcaptionskip}{-1cm}
	\label{sampletime}
\end{figure}

Fig.~\ref{sampletime} illustrates the NMSE performance versus the number of samples at SINR=-5dB. From Fig.~\ref{sampletime}, it can be seen that all  methods gradually increase in terms of NMSE as the number of samples increases. When the number of samples is small-scale,  the proposed PCA-EVD and JD perform better than SCM and SVD. In particular, the proposed JD shows an excellent NMSE performance in the small-sample case.

\begin{figure}[htb]
	\centering
	\includegraphics[height=6.5cm,width=0.9\textwidth]{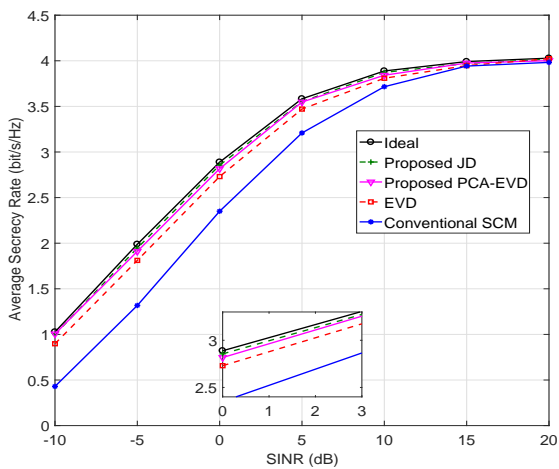}
	\caption{Curves of average SR versus SINR for proposed estimation methods utilizing ZFC RBF.}
 \setlength{\belowcaptionskip}{-1cm}
	\label{SR}
\end{figure}

Fig.~\ref{SR} plots the curves of average SR versus SINR with ZFC-RBF as a receive beamformer to reduce the jamming from Mallory, where ZFC-RBF is designed according to the estimated CMMI. The ZFC-RBF using ideal covariance matrix is constructed  as a metric. From Fig.~\ref{SR},
it can be clearly seen that the proposed JD and PCA-EVD  still perform better than EVD and SCM for -5dB$<$ SINR$<$15dB. In such an interval, they have an increasing order in SR as follows: SCM, EVD, PCA-EVD, JD. The SR of the proposed JD and PCA-EVD methods are close to the ideal one in the high SINR region.

\section{CONCLUSION}
In this paper, we have made an investigation of CMMI estimation schemes for the secure SM system with a full-duplex eavesdropper. Firstly, a rank detector was proposed using AIC as decisive criterion. Then, using the inferred rank of CMMI, two methods, PCA-EVD and JD, were proposed to improve the estimation performance from different aspects. The simulation results show that the proposed PCA-EVD performed better than  conventional SCM and EVD in NMSE and SR. The proposed JD can achieve an excellent NMSE performance, much better than other methods, in the low SINR region.  Particularly, the proposed JD  and PCA-EVD are more suitable for low-SINR and small-sample scenario.
\bibliography{bib}
\end{document}